# A practical guide for X-ray diffraction characterization of Ga(Al, In)N alloys

S.Q. Zhou*), M. F. Wu and S. D. Yaoξ)

School of Physics, Peking University, Beijing, 100871, P. R. China

## Abstract

Ga(In, Al)N alloys are used as active layer or cladding layer in light emitting diodes and laser diodes. X-ray diffraction is extensively used to evaluate the crystalline quality, the chemical composition and the residual strain in Ga(Al, In)N thin film, which directly determine the emission wavelength and the device performance. Due to the minor mismatch in lattice parameter between the Ga(Al, In)N alloy and the GaN virtual substrate, X-ray diffraction comes to a problem to separate the signal from Ga(Al, In)N alloy and GaN. In this paper, we give a detailed comparison on different diffraction planes. Generally, in order to balance the intensity and peak separation between Ga(Al, In)N alloy and GaN, (0004) and (1015) planes make the best choice for symmetric scan, and asymmetric scan, respectively.

## 1. Introduction

Driven by the commercialization of short wavelength light emitting diodes and laser diodes, the III–nitrides, aluminium nitride (AlN), gallium nitride (GaN) and indium nitride (InN), attract huge research attention [1]. They can form a continuous alloy

* E-mail address: zhousq2000@yahoo.com

ξ E-mail address: sdyao@pku.edu.cn

system (InGaN, InAlN, AlGaN, or AlInGaN) whose direct optical bandgaps range from ~0.8 eV for InN [2] and 3.4 eV for GaN to 6.2 eV for AlN. For most LED and LD devices, a thin InGaN layer or InGaN multiple quantum wells work as the active layer, while a thin AlGaN layer works as the cladding or barrier layer [3, 4]. AlInGaN quaternary alloys allow an independent control of the lattice mismatch and band offset in AlInGaN-based heterostructures [5]. The structural quality, the In/Al composition in these thin layers directly determine the emission wavelength and the performance of the device. Therefore a precise determination of the composition and the interface quality is crucial to optimize the growth procedure and the properties of the alloys. High resolution X-ray diffraction is the well-known technique to characterize the crystalline structure by measuring the lattice parameters and their spreading, consequently the composition in the compounds can be deduced [6, 7]. Usually, in the fabrication of GaN-based LED, an intermediate buffer layer of GaN (1-2 um) is grown, to act as a buffer layer and to improve crystalline quality. The InGaN or AlGaN ternary alloys are then grown onto GaN layers. This configuration arises a crucial difficulty in X-ray diffraction measurement. As shown in Fig. 1, the lattice parameter for III-Nitrides are rather close to each other [8], which results in the alloy has a even more similar lattice parameter to GaN. The minor difference in lattice parameter and the same wurtzite structure together result in the difficulty to separate the signal from Ga(Al, In)N alloys and GaN layers. In the typical X-ray diffraction spectrum, Ga(Al, In)N alloy expresses itself as a weak shoulder beside the strong peak of GaN, or even is completely overlapped by the GaN peak. In order to extract reliable information from the spectrum, it is strongly hoped that the peak of interesting layer is more separated from GaN peak, at the same time, with more intensity. In this paper, we carefully compared different diffraction planes, and gave a practical guide in X-ray diffraction measurement on Ga(Al, In)N/GaN heterostructures.

2. Experiments

The x-ray measurements were performed using high resolution x-ray diffractometer equipped with a four-crystal monochromator in Ge(220) configuration and one or two 200 µm slits before the detector. The x-ray wavelength is 1.5406 Å. The sample is mounted on a four-circle (ω, 2θ, χ, φ) geonomitor. ω and 2θ are the incident angle and diffraction angle. X is defined as the angle between the sample surface and the horizontal plane, which is defined by the incident and diffraction lines. The angle φ measures the rotation around the surface normal of the substrate. In skew symmetric diffraction geometry, the sample is tilted about the angle of the lattice plane inclination respect to the sample surface. The Ga(Al, In)N samples were grown on $Al_2O_3$(0001) by metal-organic chemical vapor deposition method. From X-ray diffraction and Rutherford backscattering/channeling measurements, all samples have very good quality.

3. Results and discussions

The basic principle of X-ray diffraction is Bragg's Law. $2d\sin\theta = n\lambda$ , where d is the lattice spacing, θ is the Bragg angle and λ is the incident X-ray wavelength. Therefore, in X-ray diffraction, the real measurement is on the variation of the lattice spacing. For symmetric (~*h* and *k=0)* Bragg reflections, only variations in lattice constant perpendicular to the sample surface can be detected. However, in the asymmetric case (measurement on the planes not parallel to the sample surface), both perpendicular and parallel lattice parameters contribute to the lattice spacing.

We investigated the symmetric case of (0002), (0004), and (0006) diffraction, and the asymmetric case of $(10\underline{1}l)$, with $l$=1 to 5, with the aim to collect the optimal spectrum. The *optimal spectrum* is supposed to possess three factors: (a) large angle separation between Ga(Al, In)N alloy and GaN, (b) small full width at half maximum (FWHM) and (c) strong diffraction intensity. In the following text, we give a practical guide to select the so-called optimal spectrum.

3.1 Symmetric scan

For symmetric scan, ω/2θ scans of (0002), (0004) and (0006) planes were studied. The intensity, the FWHMs and the peak separation are shown in Fig. 2(a) and (b). Diffraction of (0002) planes has the strongest intensity, but the smallest peak separation, while diffraction of (0006) has the biggest peak separation. Depending on In/Al content in Ga(Al, In)N alloys, (0004) planes normally make the best balance between intensity and resolution. Fig. 3 shows ω/2θ scans of (0002) and (0004) for an AlInGaN layer on GaN. The pattern of (0004) allows more reliable determination on the lattice parameter in the AlInGaN layer.

3.2 Asymmetric scan

The skew symmetric mode is used to do the asymmetric scan. ω/2θ scans of $(10\underline{1}1)$, $(10\underline{1}2)$, $(10\underline{1}3)$, $(10\underline{1}4)$ and $(10\underline{1}5)$ planes were compared in Fig. 4. The diffraction of $(10\underline{1}5)$ has the smallest FWHM (full width at half maximum), very strong intensity and biggest peak separation. Diffractions of $(10\underline{1}1)$ and $(10\underline{1}3)$ have strongest intensities, but have much bigger FWHMs, what’s more, the peak separation is smaller than that of $(10\underline{1}5)$. So $(10\underline{1}5)$ planes make the best choice for in-plane lattice constant determination. Fig. 5 compares the ω/2θ scans of an AlInGaN layer and obviously proves the optimal spectrum of (10$\underline{1}$5).

## 4. Conclusion

In a conclusion, in X-ray diffraction measurement on Ga(Al, In)N alloys, the diffraction of (0004) gives a better balance between intensity and resolution depending on the composition variation of the alloys. In asymmetric case, the diffraction of (10$\underline{1}$5) definitely makes the best choice to get the strong intensity and best resolution. This rule can also be applied to the X-ray diffraction characterization of ion-implanted and MeV ion bombarded crystalline thin film. In that case, the damaged layer has a slight elongation in the perpendicular direction [10], which comes to a similar problem of InGaN alloys on GaN.

## Acknowledgement

The authors thank Dr. Liu Jianping for providing GaN samples. This work was supported by the Key Laboratory of Heavy Ion Physics, Ministry of Education, China, and by the National Natural Science Foundation of China under grant NO. 10375004.

## Reference

[1] S. Nakamura and G. Fasol, *The Blue Laser Diode-GaN Based Light Emitter and Lasers* (Springer, Berlin, 1997).

[2] K.M. Yu, Z. Liliental-Weber, W. Walukiewicz, W. Shan, J.W. Ager III, S.X. Li, R.E. Jones, E.E. Haller,H. Lu, and W.J. Schaff, Appl. Phys. Lett. **86**, 071910-1 (2005).

[3] S. Nakamura, T. Mukai, and M. Senoh, Appl. Phys. Lett. **64**, 1687, (1994)

[4] S. Nakamura Semicond. Sci. Technol. **14** (1999) R27–R40.

[5] M. Asif Khan, J.W. Yang, G. Simin, R. Gaska, M.S. Shur,H.-C. zur Loye, G. Tamulaitis, A. Zukauskas, Appl. Phys. Lett. **76** (2000) 1161.

[6] M. Schuster, P.O. Gervais, B. Jobst, W. Hösler, R. Averbeck, H. Riechert, A. Iberl, and R. Stömmer, J. Phys. D: Appl. Phys. **32** (1999) A56.

[7] R. Singh, D. Doppalapudi, T. D. Moustakas, and L. T. Romano, Appl. Phys. Lett. **70**, 1089, (1996)

[8] Data from the website of Ioffe Institute, http://www.ioffe.rssi.ru/SVA/NSM/.

[9] R. Singh, D. Doppalapudi, T. D. Moustakas, and L. T. Romano, Appl. Phys. Lett. **70**, 1089 (1997).

[10] C. Liu, B. Mensching, K. Volz, and B. Rauschenbach, Appl. Phys. Lett. **71**, 2313 (1997).

Fig captions.

Fig. 1. Lattice parameters of III-Nitrides and their alloys. The lattice parameters of alloys are calculated from Vegard's law [9].

Fig. 2. (a) XRD intensity of GaN(000*l*); (b) FWHM of $\omega/2\theta$-scan of GaN(000*l*) and peak separation between $In_{0.1}Ga_{0.9}N$ and GaN, with $l$=2, 4, and 6. Lines are guides for eyes.

Fig. 3 Comparison of $\omega/2\theta$-scans of sample $Al_{0.03}In_{0.03}Ga_{0.94}N$ (0002) and (0004), spectrum of (0004) shows less overlapping.

Fig. 4. (a) XRD intensity of GaN($10\underline{1}l$), (b) FWHM of $\omega/2\theta$-scan of GaN($10\underline{1}l$) and peak separation between $In_{0.1}Ga_{0.9}N$ and GaN, with $l$=1, 2, 3, 4, and 5. Lines are guides for eyes.

Fig. 5. Comparison of $\omega/2\theta$-scans of sample $Al_{0.35}In_{0.02}Ga_{0.63}N$ ($10\underline{1}4$) and ($10\underline{1}5$), spectrum of ($10\underline{1}5$) shows both stronger intensity and less overlapping.

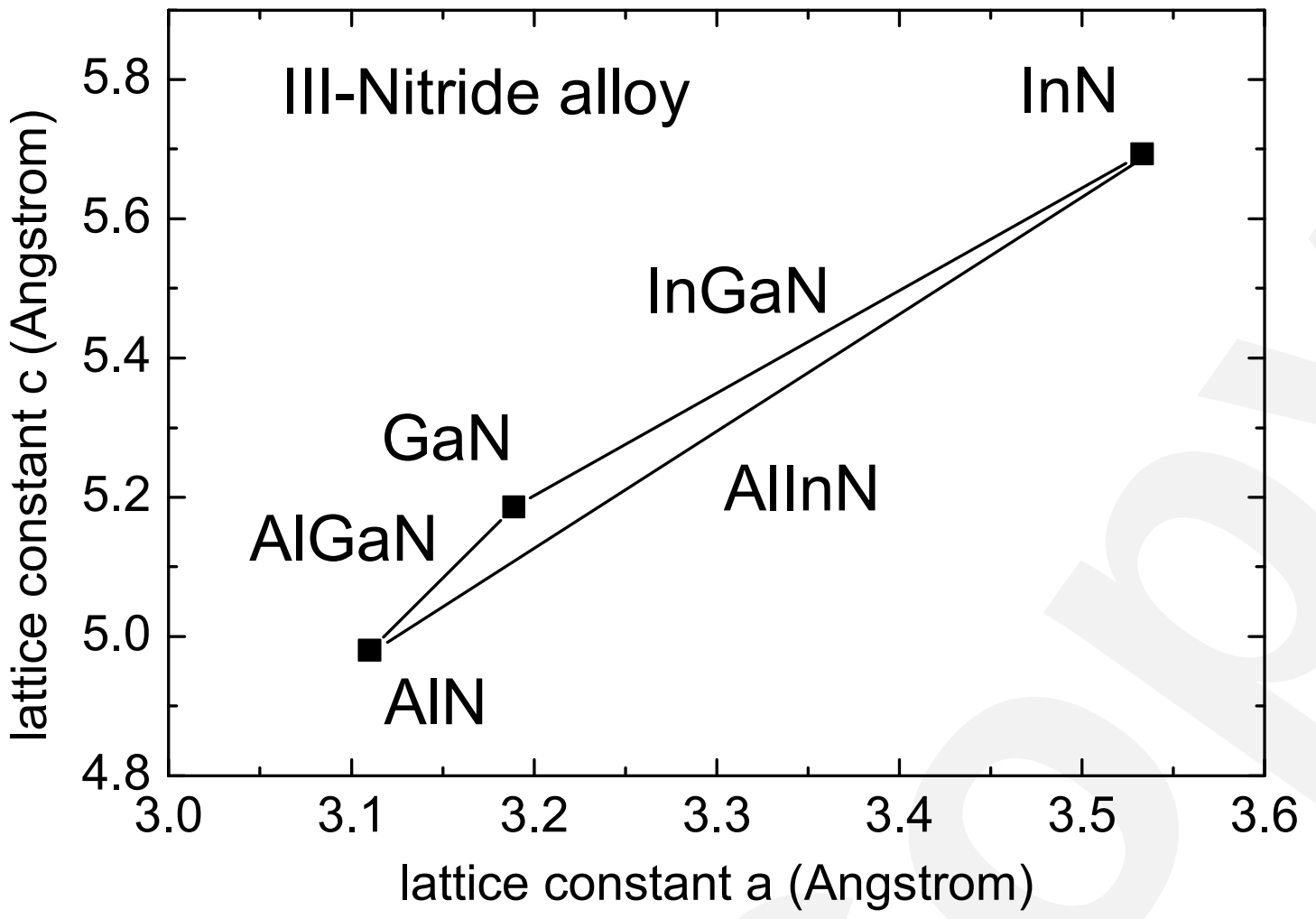

Fig. 1. Lattice parameters of III-Nitrides and their alloys. The lattice parameters of alloys are calculated from Vegard's law [9].

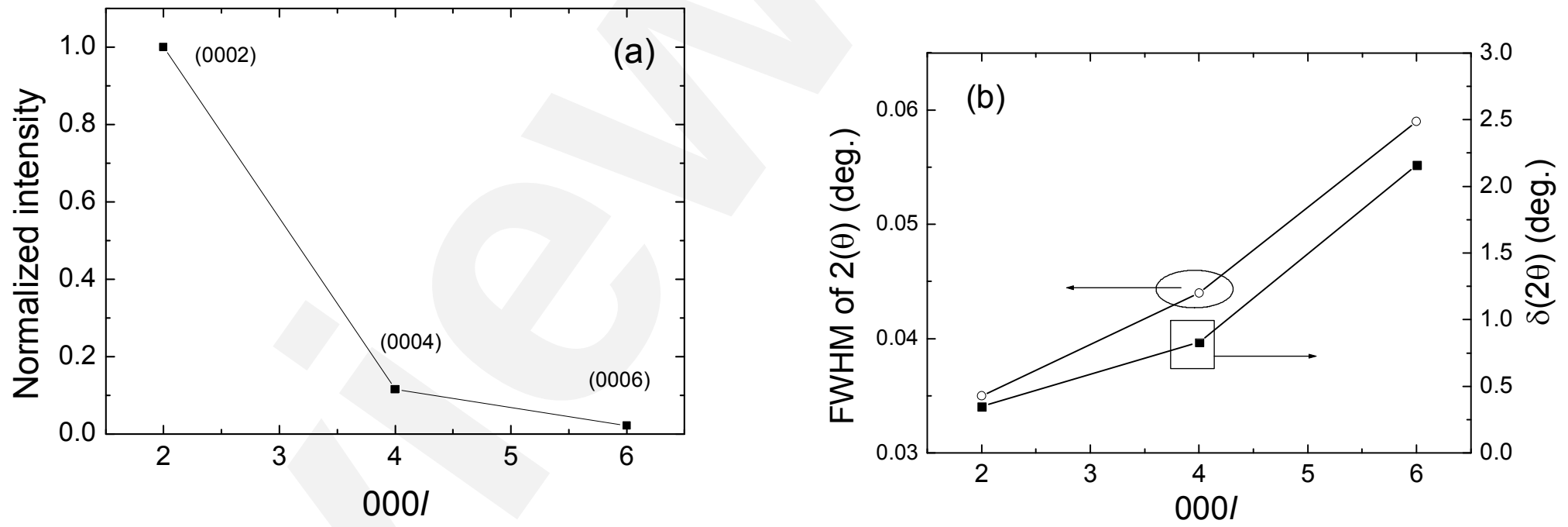

Fig. 2. (a) XRD intensity of GaN(000*l*); (b) FWHM of ω/2θ-scan of GaN(000*l*) and peak separation between $In_{0.1}Ga_{0.9}N$ and GaN, with *l*=2, 4, and 6. Lines are guiders for eyes.

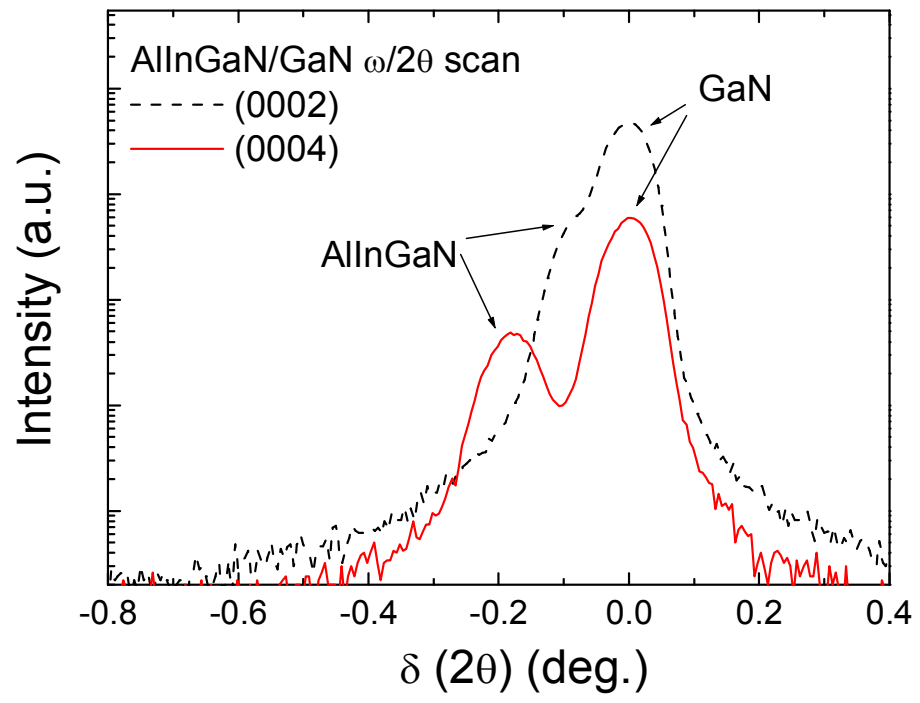

Fig. 3 Comparison of ω/2θ-scans of sample $Al_{0.03}In_{0.03}Ga_{0.94}N$ (0002) and (0004), spectrum of (0004) shows less overlapping.

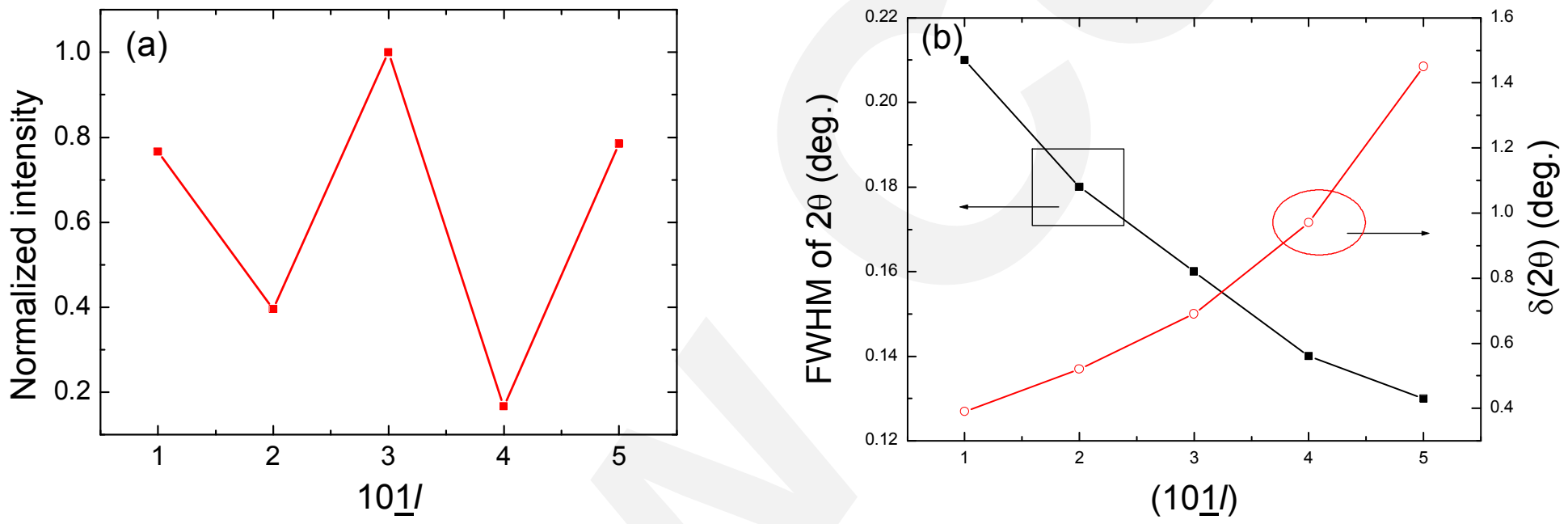

Fig. 4. (a) XRD intensity of GaN(10$\underline{1}$*l*), (b) FWHM of ω/2θ-scan of GaN(10$\underline{1}$*l*) and peak separation between $In_{0.1}Ga_{0.9}N$ and GaN, with *l*=1, 2, 3, 4, and 5. Lines are guiders for eyes.

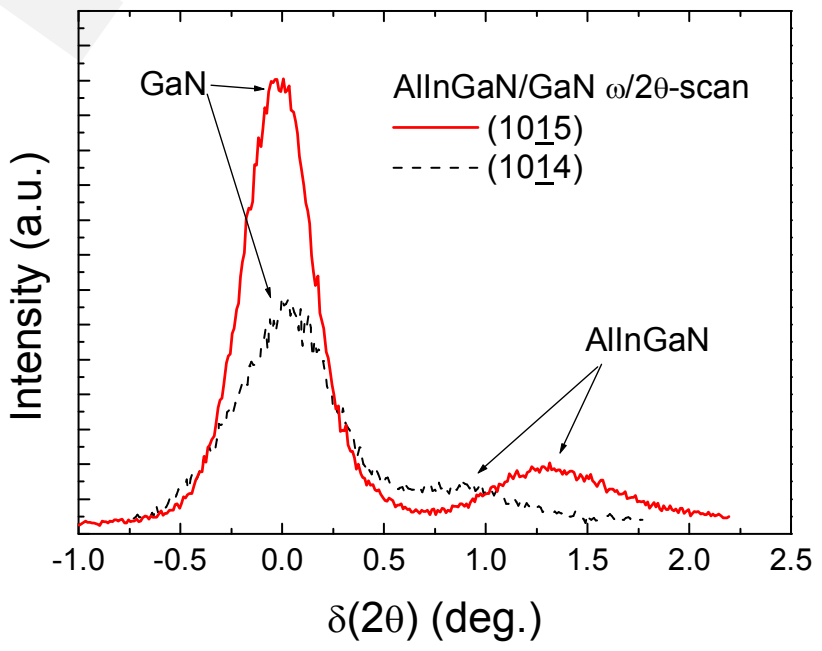

Fig. 5. Comparison of ω/2θ-scans of sample $Al_{0.35}In_{0.02}Ga_{0.63}N$ (10$\underline{1}$4) and (10$\underline{1}$5), spectrum of (10$\underline{1}$5) shows both stronger intensity and less overlapping.